\documentclass[
 reprint,
 superscriptaddress,
 nofootinbib,
 amsmath,amssymb,
 aps,
 prl,
]{revtex4-2}

\usepackage{graphicx}
\usepackage[usenames,dvipsnames]{color}
\usepackage{multirow}

\usepackage[
        range-phrase=--,
        range-units=single,
        retain-explicit-plus=true,
        retain-unity-mantissa=false,
    ]{siunitx}

\usepackage[dvipsnames,x11names]{xcolor}
\usepackage[
        colorlinks=true,
        citecolor=RoyalBlue,
        linkcolor=BrickRed,
        urlcolor=DarkGreen,
    ]{hyperref}
\usepackage{url}
\usepackage{amsmath}
\newcolumntype{P}[1]{>{\centering\arraybackslash}p{#1}}

\usepackage{soul}
\usepackage{comment}
\usepackage{orcidlink}

\definecolor{spring}{rgb}{0.7,0.9,0.7}
\definecolor{brick}{rgb}{0.7,0.2,0.1}
\definecolor{redHL}{rgb}{1.0,0.7,0.7}
\definecolor{blueHL}{rgb}{0.7,0.7,1.0}
\definecolor{ElectricBlue}{rgb}{0.49, 0.976, 1.0}
\definecolor{blueT}{rgb}{0.039, 0.729, 0.71}
\definecolor{DarkGreen}{rgb}{0, 0, 0}
\definecolor{blue}{rgb}{0.066, 0.310, 0.984}
\definecolor{black}{rgb}{0, 0, 0}

\usepackage[capitalise]{cleveref}

\newcommand{\K}{\mathcal{K}(\Omega)}
\newcommand{\hsql}{h_{\text{SQL}}(\Omega)}
\newcommand{\arm}{\text{arm}}
\newcommand{\sql}{37}

\begin{document}

\preprint{APS/123-QE}

\title{LIGO operates with quantum noise below the Standard Quantum Limit}

\author{W.~Jia\,\orcidlink{0000-0002-5119-6328}}  
\thanks{wenxuanj@mit.edu}
\affiliation{LIGO Laboratory, Massachusetts Institute of Technology, Cambridge, MA 02139, USA}

\author{V.~Xu\,\orcidlink{0000-0002-3020-3293}}
\thanks{victoriaa.xu@ligo.org}
\affiliation{LIGO Laboratory, Massachusetts Institute of Technology, Cambridge, MA 02139, USA}

\author{K.~Kuns\,\orcidlink{0000-0003-0630-3902}}
\affiliation{LIGO Laboratory, Massachusetts Institute of Technology, Cambridge, MA 02139, USA}

\author{M.~Nakano\,\orcidlink{0000-0001-7703-0169}}
\affiliation{LIGO Livingston Observatory, Livingston, LA 70754, USA}

\author{L.~Barsotti\,\orcidlink{0000-0001-9819-2562}}
\affiliation{LIGO Laboratory, Massachusetts Institute of Technology, Cambridge, MA 02139, USA}

\author{M.~Evans\,\orcidlink{0000-0001-8459-4499}}
\affiliation{LIGO Laboratory, Massachusetts Institute of Technology, Cambridge, MA 02139, USA}

\author{N.~Mavalvala\,\orcidlink{0000-0003-0219-9706}}
\affiliation{LIGO Laboratory, Massachusetts Institute of Technology, Cambridge, MA 02139, USA}


\author{\\R.~Abbott}
\affiliation{LIGO Laboratory, California Institute of Technology, Pasadena, CA 91125, USA}
\author{I.~Abouelfettouh}
\affiliation{LIGO Hanford Observatory, Richland, WA 99352, USA}
\author{R.~X.~Adhikari\,\orcidlink{0000-0002-5731-5076}}
\affiliation{LIGO Laboratory, California Institute of Technology, Pasadena, CA 91125, USA}
% ...

\begin{abstract} 
Precision measurements of space and time, like those made by the detectors of the Laser Interferometer Gravitational-wave Observatory (LIGO), are often confronted with fundamental limitations imposed by quantum mechanics. The Heisenberg uncertainty principle dictates that the position and momentum of an object cannot both be precisely measured, giving rise to an apparent limitation called the Standard Quantum Limit (SQL). 
Reducing quantum noise below the SQL in gravitational-wave detectors, where photons are used to continuously measure the positions of freely falling mirrors, has been an active area of research for decades. 
Here we show how the LIGO A+ upgrade reduced the detectors' quantum noise below the SQL by up to \SI{3}{dB} while achieving a broadband sensitivity improvement, more than two decades after this possibility was first presented.
\end{abstract}

\maketitle

\textbf{Introduction --} One of the most profound consequences of quantum mechanics is the Heisenberg uncertainty principle, which posits that the product of the measurement noises
of conjugate observables (i.e, position and momentum) cannot be less than $\hbar/2$. 
Measuring the position $x$ of an object with an uncertainty $\Delta x$ inevitably perturbs its momentum by $\Delta p \geq \hbar/(2 \Delta x)$. 
After a time $\tau$, the massive object $m$ will freely evolve with additional position uncertainty $\Delta x'$ from the momentum perturbation $\Delta x' = \tau \Delta p /m = \hbar \tau/(2m\Delta x)$. An extremely precise measurement ($\Delta x \rightarrow 0$) will make the next position measurement totally unpredictable ($\Delta x' \rightarrow \infty$) due to quantum back action~\cite{braginskyS80QuantumNondemolition}. The minimal possible uncertainty can be achieved with $\Delta x = \Delta x' = \sqrt{\hbar \tau/(2m)}$, which is known as the Standard Quantum Limit (SQL)~\cite{braginsky92QuantumMeasurement}. While the SQL applies to measurements of microscopic particles, it is also a limiting factor for the measurements made by the LIGO interferometric detectors, which probe attometer-scale displacements of macroscopic mirrors~\cite{aLIGO}. 

In the 1980s, it was suggested that the SQL could be surpassed by introducing quantum correlations between the interferometer's laser light and the mirrors~\cite{unruhQOEGaMT83QuantumNoise, unruhQOEGaMT83ReadoutState, yuenPRL83ContractiveStates}. 
In the early 2000s, proposed designs to convert the LIGO into ``quantum nondemolition interferometers''\footnote{The concept of quantum nondemolition measurements was introduced to describe measurements with quantum noise below the SQL~\cite{braginskyS80QuantumNondemolition, braginsky92QuantumMeasurement, braginskyRMP96QuantumNondemolition, braginskiiSPU75QuantummechanicalLimitations}. We note that this definition, in certain scenarios, differs from what is adopted in other fields of physics, where a quantum nondemolition measurement implies that there is no quantum back action on the measured observable~\cite{QND_Haroche, QND_NIST}.} emerged~\cite{kimblePRD01ConversionConventional}; one approach, referred to as ``squeezed-input" interferometer, suggested that it is possible to break the SQL at a particular frequency by injecting a non-classical state of light, known as squeezed vacuum state, to the LIGO interferometer. Furthermore, the addition of a detuned Fabry-P\'erot ``filter cavity" would produce a frequency-dependent phase shift on the squeezed vacuum states reflected from it and therefore enable a broadband quantum enhancement below the SQL. 

In the first proof-of-principle demonstration in 2020, we injected squeezed vacuum states in LIGO to demonstrate quantum correlations and surpass the SQL in a narrow frequency region of the detection band (\SIrange{30}{50}{Hz})~\cite{Haocun_nature}. However, since this was achieved without a filter cavity, it led to a quantum noise increase at frequencies outside the sub-SQL dip and an overall decrease of the astrophysical sensitivity, as predicted in~\cite{kimblePRD01ConversionConventional}.

As part of the LIGO A+ upgrade that started in 2022, a 300-m long filter cavity was added to both LIGO Livingston (L1) and Hanford (H1) interferometers to achieve broadband reduction of quantum noise~\cite{O4FDS}. Here we present the first modeling and analysis of quantum noise in the LIGO interferometer operating with a filter cavity. We show that LIGO's quantum noise surpasses the SQL by up to \SI{3}{dB} between \SI{35}{Hz} and \SI{75}{Hz} in astrophysical operation, with enhanced sensitivity in most of the detection band, thereby realizing the goal first set out over two decades ago~\cite{kimblePRD01ConversionConventional}.


\textbf{Theory --} 
Gravitational-wave modulations of spacetime are quantified by strain $h$. The LIGO gravitational-wave interferometer converts these modulations into a measurable differential displacement between two pairs of suspended mirrors. The dimensionless gravitational-wave strain and the interferometer differential displacement $\Delta x$ are related by
$h = \Delta x / L_{\arm}$, where $L_{\arm} = \SI{4}{km}$ is the length of the each interferometer arm.
At relevant measurement frequencies, the interferometer mirrors move freely, and the SQL for these mirrors can be expressed in units of gravitational-wave strain noise amplitude spectral density as
\begin{equation} \label{eq: SQL}
    \hsql = \frac{\Delta x_\text{SQL} (\Omega)}{L_{\arm}} = \sqrt{\frac{2\hbar}{(m/4)\Omega^2}} \frac{1}{L_{\arm}} 
\end{equation}
where $\hbar$ is the reduced Planck's constant, and $\Omega$ is the measurement frequency. Notably, this limit depends on the mass of the object rather than the number of photons used to probe the object (i.e., the laser power). In LIGO, the mass of the object is the reduced mass ($m/4$) of the differential motion of each pair of arm cavity mirrors with mass $m = \SI{40}{kg}$

$$h_\text{SQL}^{40 \text{kg}} (\Omega) \approx 1.8\times10^{-24} \left(\frac{2\pi \times \SI{100}{Hz}}{\Omega}\right) \frac{1}{\sqrt{\text{Hz}}}.$$ 

In~\cite{O4FDS}, we presented a simplified model of quantum noise in the LIGO interferometers. In the ideal lossless case, the power spectral density (PSD) of quantum noise can be expressed in units of strain as: 
\begin{equation} \label{eq:unsqz_noise_model}
    S(\Omega) = \frac{h_\text{SQL}^2(\Omega)}{2} \left( \K + \frac{1}{\K} \right)
\end{equation}
where the first term represents noise due to quantum back action and the second term represents imprecision noise. The optomechanical coupling strength $\K$ increases with the circulating laser power in arm cavities $P_{\arm}$ as
\begin{equation} \label{eq: kappa}
    \K = \frac{16 k_0 P_{\arm}}{m \gamma_0 L_{\arm}} \frac{1}{\Omega^2} \left(1+\frac{\Omega^2}{\gamma_0^2}\right)^{-1}.
\end{equation}
where $k_0=2\pi/(\SI{1064}{nm})$ is the laser wavenumber, and $\gamma_0 \approx 2\pi\times\SI{450}{Hz}$ is the detector's signal bandwidth.

At frequencies below \SI{100}{Hz}, measurement back action dominates due to the strong opto-mechanical coupling ($\K \gg 1$). At frequencies above $\gamma_0$, the opto-mechanical coupling is weak ($\K \ll 1$), and the measurement imprecision from photon shot noise dominates. 
These two forms of quantum noise contribute equally to the total quantum noise at the SQL frequency $\Omega_\text{SQL}$, defined by $\mathcal{K}(\Omega_\text{SQL}) = 1$. Note $\Omega_\text{SQL}$ scales with the square root of the laser power; for a circulating power of $P_{\arm} = \SI{260}{kW}$, $\Omega_\text{SQL} = 2\pi \times \SI{\sql}{Hz}$.

Together, these two forms of quantum noise enforce the so-called SQL for displacement sensing (\cref{eq:unsqz_noise_model}), which arises from the use of uncorrelated photons to probe mirror positions. 
\cref{eq:unsqz_noise_model} enforces the SQL because it is an incoherent superposition of quantum back action and imprecision noise. In the presence of quantum correlations between light and mirrors, \cref{eq:unsqz_noise_model} no longer holds, allowing the SQL to be surpassed.

Squeezed vacuum is a non-classical state of light which uses quantum correlations between photon pairs to reduce one form of quantum noise (e.g. imprecision noise) at the expense of the other (e.g. quantum back action noise), in the way allowed by the Heisenberg uncertainty principle~\cite{SCHNABEL20171}. 
The injection of squeezed vacuum into the output port of an interferometer~\cite{cavesPRD81QuantummechanicalNoise} modifies its quantum noise relative to \cref{eq:unsqz_noise_model} to produce
\begin{equation} \label{eq: model}
    S_{\text{SQZ}}(\Omega) = S(\Omega) \left[ e^{-2r}\cos^2(\phi - \theta(\Omega)) + e^{2r}\sin^2(\phi - \theta(\Omega)) \right]
\end{equation}
where $e^{-2r}$ is the factor by which the injected quantum noise is squeezed relative to vacuum noise, 
$\phi$ is the relative phase between the input squeezed field and the interferometer field (i.e. the ``squeeze angle''), and $\theta(\Omega) = \tan^{-1}\K$ is the squeeze angle rotation due to the optomechanical response of the interferometer.

Frequency-dependent squeezed states, where the squeeze angle varies as a function of frequency $\phi\to\phi(\Omega)$, can be prepared by reflecting the frequency-independent squeezed state from a detuned and overcoupled Fabry-P\'erot cavity ~\cite{mccullerPRL20FrequencyDependentSqueezing, zhaoPRL20FrequencyDependentSqueezed, O4FDS, virgoFDS}. When the filter cavity linewidth is well-matched to $\Omega_\text{SQL}$, it imparts the phase rotation $\phi(\Omega) \approx \theta(\Omega) = \tan^{-1}\K$ upon the reflected squeezed vacuum and enables quantum noise reduction of $e^{-2r}$ at all frequencies~\cite{whittlePRD20OptimalDetuning}:
\begin{equation} \label{eq: FDSQZ}
    S_{\text{FDSQZ}}(\Omega) = \frac{h_\text{SQL}^2(\Omega)}{2} \left( \K + \frac{1}{\K} \right)e^{-2r}.
\end{equation}
In particular, around $\Omega_\text{SQL}$, quantum noise is reduced below the SQL by a factor of $e^{-2r}$
\begin{equation}
    S_{\text{FDSQZ}}(\Omega_\text{SQL}) = h_\text{SQL}^2(\Omega_\text{SQL})e^{-2r}.
\end{equation}

\begin{figure}[t!]
\begin{center}
    \includegraphics[width=1.0\linewidth]{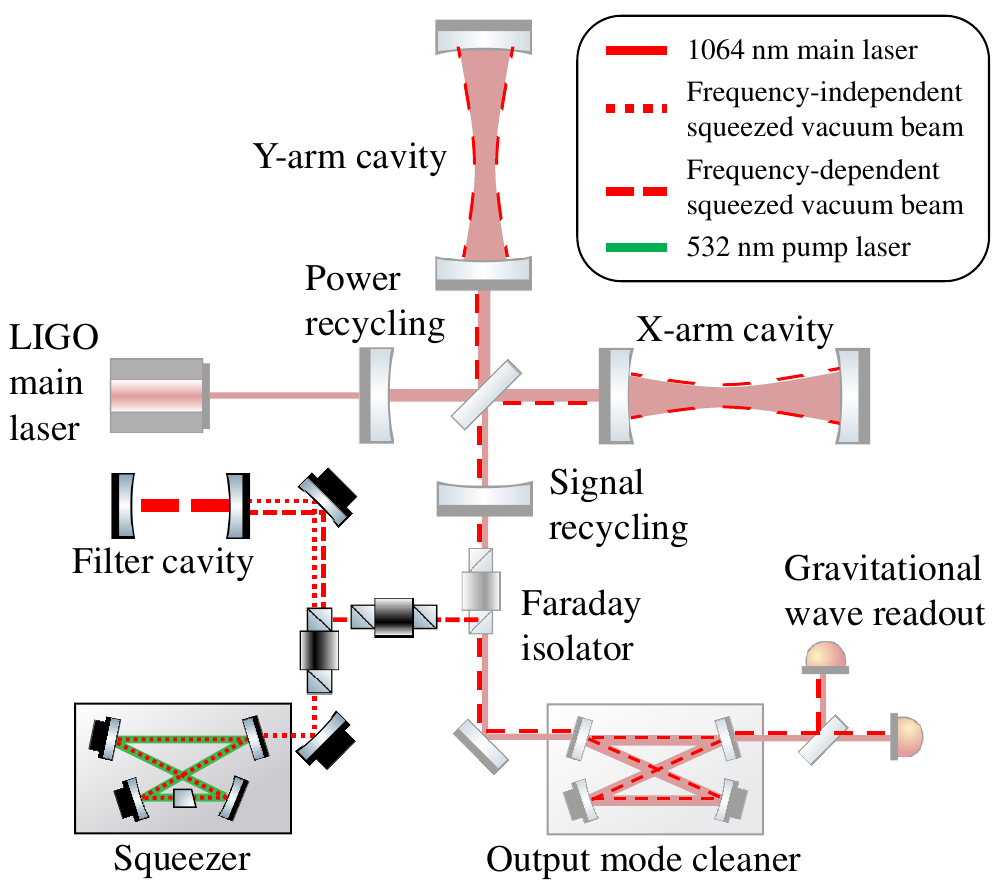}
    \caption{Simplified schematic of the LIGO A+ interferometer as of the fourth astrophysical observing run (O4) started in 2023. The squeezing system is shown overlaying the shaded main interferometer, which is a dual-recycled Michelson with 4-km long arm cavities. All optical components shown, except for the main laser, are suspended in ultra-high vacuum. Frequency-independent squeezed vacuum is generated by an optical parametric amplifier (``squeezer''), which consists of a nonlinear optical crystal in a dually-resonant bowtie cavity. The outgoing squeezed beam is reflected from a 300-m long filter cavity to produce frequency-dependent squeezing, injected via the Faraday isolator, and then propagated through the full LIGO interferometer.
    }
    \label{fig: LIGO}
\end{center}
\end{figure}

\begin{figure*}[!ht]
\begin{center}
    \includegraphics[width=0.9\linewidth,trim={1cm 0cm 1cm 2cm}]{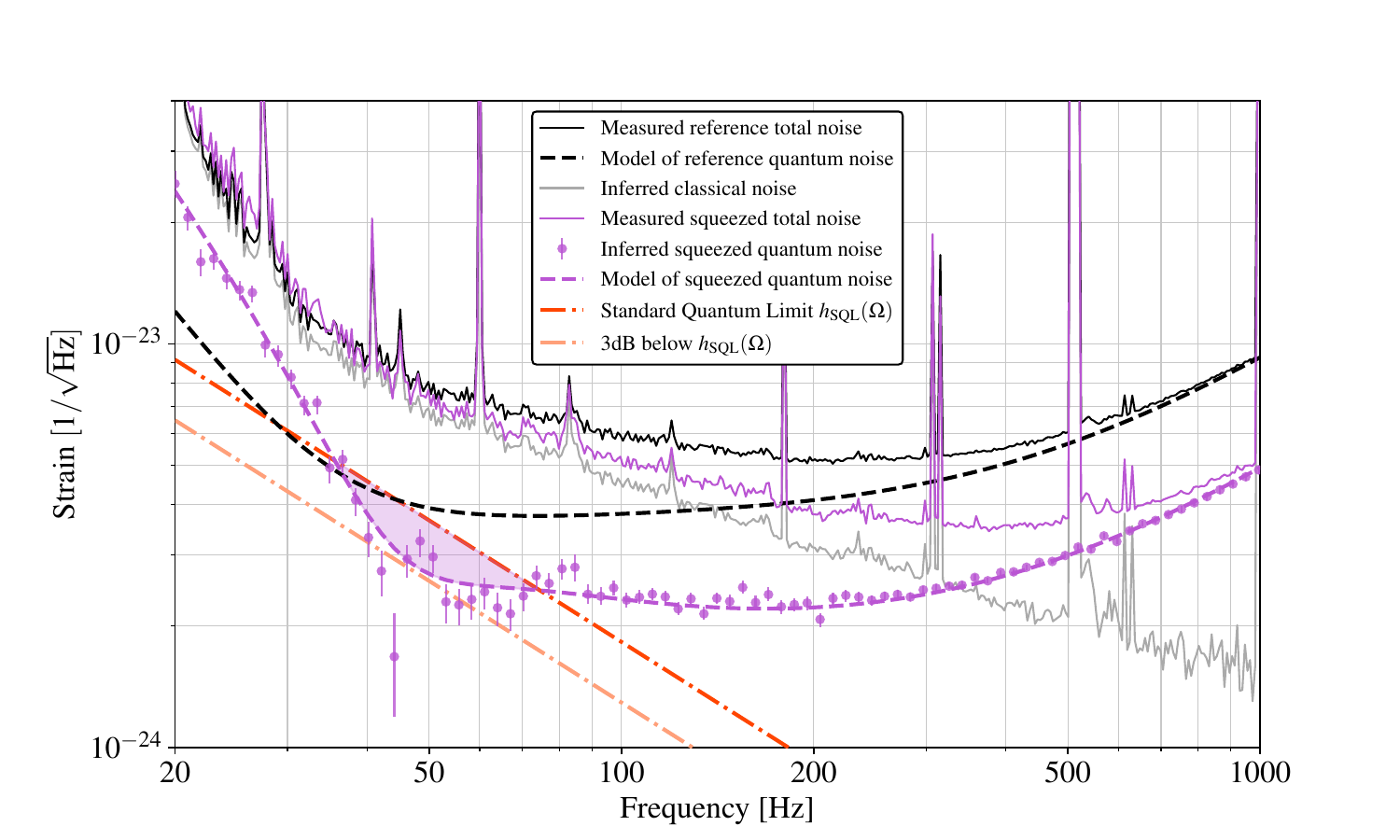}
    \caption{Strain sensitivity of the LIGO L1 interferometer. The squeezed quantum noise surpasses the standard quantum limit $\hsql$ by up to \SI{3}{dB} in the shaded region between \SIrange{35}{75}{Hz}. Error bars indicate the total 1-$\sigma$ uncertainty. This configuration is representative of the nominal detector noise during O4, demonstrating the use of quantum correlations to directly improve astrophysical sensitivity. The total detector noise spectrum is an incoherent sum of the classical and quantum noise. The unsqueezed reference total noise (solid black) is measured without squeezing injection. An unsqueezed quantum noise model (dashed black) is subtracted from the measured reference total noise to obtain an estimate of the underlying classical noise (gray). The inferred detector quantum noise with squeezing (purple dots) is obtained by subtracting the classical noise estimate (gray) from the measured squeezed total noise spectra (solid purple). The dashed purple trace shows a fitted model of frequency-dependent squeezed noise spectra, given our best knowledge of the detector and squeezer parameters.
    }
    \label{fig: moneyplot}
\end{center}
\end{figure*}

\textbf{Experimental setup --} \cref{fig: LIGO} shows a simplified diagram of the LIGO interferometer~\cite{Aaron}, which includes Fabry-P\'erot arm cavities formed by a pair of 40-kg mirrors to resonantly enhance strain sensitivity, input power recycling to increase the circulating laser power (and thus $\K$), and output signal extraction to broaden the detection bandwidth.
Components of the squeezing system, comprising the squeezed vacuum source (``squeezer") and the filter cavity, are highlighted in the figure.

Squeezed vacuum is injected at the output port of the interferometer to reduce quantum noise~\cite{Barsotti_2019}. The LIGO squeezer generates frequency-independent squeezed vacuum via spontaneous parametric down-conversion in a bowtie optical parametric amplifier cavity containing a nonlinear PPKTP crystal~\cite{Maggie, Chua:11}. 
As described in~\cite{kimblePRD01ConversionConventional, O4FDS}, the 300-m filter cavity is controlled on the resonance at a detuned frequency with respect to the carrier frequency of the main laser, thus producing frequency-dependent squeezing ($\phi \rightarrow \phi(\Omega)$) before injection into the interferometer.

\textbf{Results --} 
\cref{fig: moneyplot} shows the first detailed quantum noise analysis of the LIGO L1 detector operating with frequency-dependent squeezing. Beyond our previous work~\cite{O4FDS} that shows only the total detector noise reduction with frequency-dependent squeezing, here we demonstrate quantum noise that surpasses the SQL between \SIrange{35}{75}{Hz}, by as much as \SI{3}{dB} near \SI{50}{Hz}, as highlighted in the purple shaded regions. While a complete analysis was done only for the L1 interferometer data, qualitatively similar results were observed in H1.

Accurate estimation of squeezed quantum noise below \SI{100}{Hz} is complicated by the presence of non-quantum (``classical") noises that are a factor of 2 higher in amplitude. In this work, we performed further measurements and extensive analysis to accurately infer the squeezed quantum noise from total noise measurements. 

\begin{figure}[ht!]
\begin{center}
    \includegraphics[width=\linewidth]{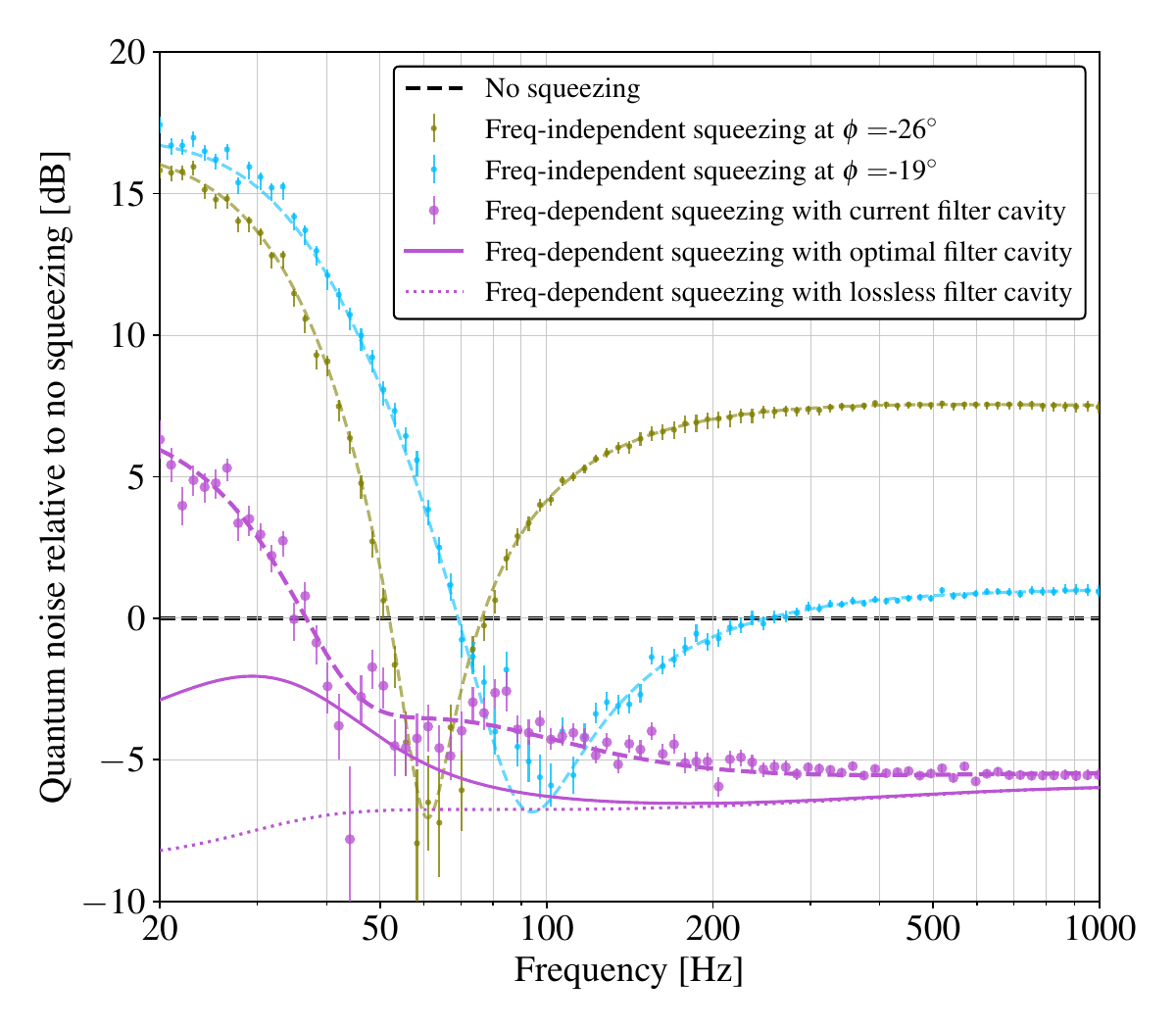}
    \caption{
    Quantum noise reduction in units of decibels. Dots show the inferred quantum noise from measurements of the total detector noise in various configurations. Dashed traces are the quantum noise models. The input filter cavity rotates the injected squeezing angle $\phi$ as a function of frequency to produce frequency-dependent squeezing ($\phi \to \phi(\Omega)$)~\cite{O4FDS}. Blue and olive traces show the inferred quantum noise with frequency-independent squeezing injected at two different $\phi$. They outline the minimum quantum noise achievable at particular frequencies, given detector losses. The three purple traces show the quantum noise with (i) frequency-dependent squeezing using the current filter cavity (dashed purple, same as \cref{fig: moneyplot}), (ii) an optimal filter cavity (solid purple) with a cavity linewidth well-matched to the current circulating laser power and 60-ppm round-trip optical loss, and (iii) a lossless optimal filter cavity (dotted purple). 
    }
    \label{fig: QND_FIS}
\end{center}
\end{figure}

There are two steps to inferring quantum noise. First, we infer the classical noise (gray) by subtracting an unsqueezed quantum noise model (dashed black) from measurements of the total unsqueezed detector noise (solid black). An accurate model of unsqueezed quantum noise is crucial to determine classical noise from subtraction. The model is known to have degenerate parameters. For example, the circulating power and optical loss in the readout path affect the imprecision noise in the same way phenomenologically. To constrain the parameter space, we experimentally set a few constant squeezing angles $\phi$ and find a set of interferometer parameters that accurately models the measured noise for each $\phi$, since the quantum noise $S_{\text{SQZ}}$ heavily depends on $\phi$ (\cref{eq: model}). We perform a Markov Chain Monte Carlo inference to find a set of parameters that make a common fit to all 11 different squeeze angle datasets (with a subset of these data shown in \cref{fig: QND_FIS}). These parameters include key experimental non-idealities such as squeezing phase noise, optical loss and mode-mismatches across the interferometer, as described in~\cite{buonannoPRD01QuantumNoise,DnD,Lee_LIGO_response}. 
Second, we subtract this classical noise estimate from subsequent measurements of the total detector noise with squeezing (purple) to infer the squeezed quantum noise (purple dots), representing our measure of $\sqrt{S_\text{SQZ}(\Omega)}$ from \cref{eq: model}. Detailed discussion of inferred parameters and model residuals are contained in the Supplemental Material. 

The two-step noise subtraction process assumes that classical noise remains identical across unsqueezed and squeezed modes of operation. Variations in classical noise between these modes will then appear as estimation uncertainties. Here we use the same uncertainty propagation methods as~\cite{Haocun_nature} to estimate the total error budget, including statistical uncertainties from detector noise PSD estimation and non-stationary classical noise, and the systematic uncertainties from calibration and residual model errors.

Statistical uncertainties limit our estimation of low frequency quantum noise. This includes uncertainties from PSD estimation (requiring long averaging times) and non-stationary classical noise (requiring technical detector improvements)~\cite{detcharO3}. To reduce the statistical uncertainty, the total noise measurements in \cref{fig: moneyplot} were obtained by averaging the detector noise over \SIrange{0.5}{1}{hour} in each configuration, and by alternating between unsqueezed and squeezed configurations to control for time variations of the classical noise. 
We find that differences in the classical noise between segments (non-stationarity) was comparable to the total uncertainty from one hour of PSD estimation with optimal frequency binning.

The main systematic uncertainty arises from the real-time calibration process, where we apply a known force to the mirror to actively modulate the strain and measure the instrument's response~\cite{calib_paper, Pcal}. 
For the data shown here, the systematic uncertainties are less than $5\%$. Full derivations of total uncertainty budget can be found in Supplemental Materials.

\cref{fig: QND_FIS} shows L1 measurements of the inferred quantum noise with frequency-dependent squeezing (purple traces) and frequency-independent squeezing at two injected squeeze angles $\phi$, in decibels of quantum noise reduction compared to no squeezing (i.e., $20 \log_{10} \left[\sqrt{S_\text{SQZ}(\Omega)/S(\Omega)}\right]$). 
Dashed traces show numerical quantum noise models that include best-fit experimental parameters for the full interferometer, squeezer, and filter cavity. 
Strong agreements between model curves (dashed traces) and measured spectra (dots with error bars) support the unsqueezed quantum noise model used for subtraction and experimental parameters for the squeezer. 
This model is then extended to include the filter cavity parameters, initially described in~\cite{O4FDS}. The quantum noise models with frequency-dependent squeezing (dashed purple curve) agree well with the inferred quantum noise spectra (purple dots).

While the current frequency-dependent squeezing configuration achieves quantum-noise suppression above \SI{35}{Hz} (see the dashed purple ``current FC'' curve in \cref{fig: QND_FIS}), frequency-independent squeezing models and measurements all suggest that an optimal filter cavity would yield significantly greater quantum noise reduction at astrophysically-important low frequencies (solid purple ``optimal FC'' curve). 
The discrepancy between the current and optimal filter cavity arises from the mismatch between the current SQL frequency and the filter cavity linewidth. In the lossless case, the optimal filter cavity would have an equal half-width-half-maximum linewidth $\gamma_{\text{FC}}$ and detuning both determined by the SQL frequency, $\gamma_{\text{FC}}= \Omega_{\text{SQL}}/ \sqrt{2}$~\cite{whittlePRD20OptimalDetuning}. The current filter cavity was designed to have $\gamma_{\text{FC}} = 2\pi \times 42$ Hz, using an input coupler power transmissivity of $T_\text{in} \approx 1000$ ppm~\cite{O4FDS} and assuming 60 ppm optical loss, to approximately match $\Omega_{\text{SQL}} =  \sqrt{2}\gamma_{\text{FC}} = 2\pi \times 59$ Hz. However, the current SQL frequency is at $\Omega_\text{SQL} = 2\pi\times 37$ Hz. Since $\Omega_{\text{SQL}}$ is proportional to the square root of arm power as in \cref{eq: kappa}, the optimal filter cavity curve in \cref{fig: QND_FIS} could be approached by either reducing the filter cavity linewidth (reducing $T_\text{in}$, solid purple), or increasing the current arm power from \SI{260}{kW} to \SI{500}{kW}, as shown in \cref{fig: future_FC}. This is because a higher circulating laser power couples back action into the measurement over a larger bandwidth, requiring a higher bandwidth filter cavity to compensate.

Compared to frequency-independent squeezing spectra (blue and olive traces), the lossless optimal filter cavity rotates the injected squeezing angle as a function of frequency, $\phi\to\phi(\Omega) \approx \tan^{-1}\K$, to approach the minimum quantum noise at all frequencies simultaneously (dotted purple). Ideally, frequency-dependent squeezing is a single configuration that reaches the envelope of minimal quantum noises achievable by all frequency-independent spectra.

\textbf{Conclusions --} With frequency-dependent squeezing, the LIGO A+ detectors now operate with quantum-limited sensitivity surpassing the SQL, as envisioned for the first time over two decades ago~\cite{kimblePRD01ConversionConventional}.  
The methods described here enabled us to accurately model quantum noise through the complex optical systems of the LIGO interferometers, with important insights that inform the next steps toward the A+ target of \SI{6}{dB} of broadband squeezing enhancement.

Concepts for future upgrades in the LIGO facilities and the next generation of gravitational-wave detectors like Cosmic Explorer~\cite{CE} and Einstein Telescope~\cite{ET} include the ambitious goal of 10 dB squeezing enhancement. Techniques and methods presented here are fundamental to achieving this goal and further enhancing the scientific potential of gravitational-wave observatories.

\section*{Acknowledgments} 
The authors would like to thank Vivishek Sudhir, Dennis Wilken, and Harald Pfeiffer for helpful discussions.
\textbf{Funding:} The authors gratefully acknowledge the support of the United States National Science Foundation (NSF) for the construction and operation of the LIGO Laboratory and Advanced LIGO as well as the Science and Technology Facilities Council (STFC) of the United Kingdom, and the Max-Planck-Society (MPS) for support of the construction of Advanced LIGO. Additional support for Advanced LIGO was provided by the Australian Research Council. The authors acknowledge the LIGO Scientific Collaboration Fellows program for additional support. LIGO was constructed by the California Institute of Technology and Massachusetts Institute of Technology with funding from the National Science Foundation, and operates under cooperative agreement PHY-2309200. Advanced LIGO was built under award PHY-18680823459. The A+ Upgrade to Advanced LIGO is supported by US NSF award PHY-1834382 and UK STFC award ST/S00246/1, with additional support from the Australian Research Council. 
\textbf{Author contributions:} L.B., M.E., and N.M. conceived and supervised the project. W.J. and M.N. optimized and configured the LIGO Livingston detector to collect the data. W.J., V.X., and K.K. analyzed the data with the model developed by L.M. and K.K.. W.J., V.X., L.B., and M.E. prepared the manuscript. 
\textbf{Competing interests:} No conflict of interest. 
\textbf{Data and materials availability:} All data are available in the manuscript or the supplementary materials.

\bibliography{QND}

\newpage
\onecolumngrid

\appendix

\section*{Supplemental Material}\label{sec:supplement}

\section{Uncertainty Analysis}

The argument that the LIGO detector operates beyond the standard quantum limit (SQL) requires a high statistical significance of the inferred quantum noise below the SQL. We follow our previous work \cite{Haocun_nature} to estimate the total uncertainty of the inferred quantum noise. The formalism is briefly summarized here. The strain noise power spectral density (PSD) of inferred quantum noise $Q(\Omega)$ is obtained from measured reference (unsqueezed) total noise $D_r(\Omega)$, measured squeezed total noise $D_s(\Omega)$, and model of the reference quantum noise $M(\Omega)$:
\begin{equation} \label{eq: Q}
    Q(\Omega) = D_s(\Omega) - (D_r(\Omega) - M_r(\Omega)) .
\end{equation}
The total uncertainty of $Q(\Omega)$ is
\begin{equation}
    \Delta Q^2(\Omega) = Q^2(\Omega) \delta G_{cal}^2(\Omega) + \left[\Delta D_s^2(\Omega) + \Delta D_r^2(\Omega) + \Delta M_r^2(\Omega) + (D_r(\Omega) - M_r(\Omega))^2 (\delta N_t^2(\Omega) + \delta N_m^2(\Omega)) \right]
\end{equation}
where
\begin{itemize}
    \item $\delta G_{cal}(\Omega)$ is the reported combined calibration error and uncertainty estimate \cite{calib_paper},
    \item $\Delta D(\Omega)$ is the statistical uncertainty due to PSD estimation,
    \item $\Delta M_r(\Omega)$ is the uncertainty of the unsqueezed reference quantum noise model, and
    \item $\delta N(\Omega)$ describes the non-stationary changes in the classical noise contributions, where $\delta N_t(\Omega)$ is time-nonstationarity and $\delta N_m(\Omega)$ is the operating mode nonstationarity between unsqueezed and squeezed operating modes.
\end{itemize}

In this paper, we follow the convention in \cite{Haocun_nature} and use $\Delta$ to describe the 1-$\sigma$ uncertainty of the variable, and use $\delta$ for the relative uncertainty $\delta D = \Delta D/D$. We plot the noise spectrum in units of amplitude spectral density (ASD) $q(\Omega) = \sqrt{Q(\Omega)}$. The relative error in ASD is
\begin{equation}
    \delta q(\Omega) = \dfrac{1}{2} \delta Q(\Omega) =  \sqrt{\dfrac{\delta G_{cal}^2(\Omega)}{4} + \dfrac{1}{4 Q^2(\Omega)} \left[\Delta D_s^2(\Omega) + \Delta D_r^2(\Omega) + \Delta M_r^2(\Omega) + C^2(\Omega) (\delta N_t^2(\Omega) + \delta N_m^2(\Omega)) \right]} .
\end{equation}

\subsection{Re-binning Power Spectral Density}

The statistical uncertainty $\delta D$ of the PSD scales inversely with the square root of the number of averages, which is proportional to the product of the duration $T$ of the time series and the frequency bin width $f$
\begin{equation} \label{eq: deltaD}
    \delta D = \frac{1}{\sqrt{ T f}} .
\end{equation}

We first take the linear FFT of the raw time series to estimate the total noise PSD. For each frequency bin, we take the median statistics to indirectly remove potential glitches in the time series, as described in our previous work \cite{Haocun_nature}. 

The linearly spaced PSD has the constant frequency bin width, for which we choose a frequency resolution of \SI{0.0625}{Hz}. To reduce the statistical uncertainty and fit the model, we re-bin the PSD into a log-spaced frequency bins. Each new frequency bin collects all the energy of the old frequency bins that falls into the bin so that the total spectral energy is conserved. The statistical uncertainty of the new PSD with log-spaced and larger bin width still follows the relation of \cref{eq: deltaD}.

The raw PSD measures the total differential displacement between the two pairs of arm cavity mirrors, which contain many peaks and resonances including harmonics of the 60-Hz power line and 500-Hz violin mechanical modes of test masses suspensions, etc. These peaks would inflate the energy of our re-binned PSD. Therefore, we remove all the known noise peaks before re-binning. 

\subsection{Non-stationarity Verification}
The stationarity uncertainty has two contributing terms: time-nonstationarity $\delta N_t (\Omega)$ that captures slow thermal drifts of the interferometer, and mode-nonstationarity $\delta N_m(\Omega)$ that contains changes introduced by different operating modes of the interferometer, namely with and without squeezing. 

To measure the unsqueezed total noise as closely as the configuration with frequency-dependent squeezing, we set up the squeezing configuration but without squeezed vacuum generated. Specifically, we leave both the squeezer and filter cavity locked on resonance but without the nonlinear parametric down-conversion process. As seen in \cref{fig: LIGO}, we only send auxiliary control sidebands to the squeezer cavity for lock acquisition \cite{O4FDS}, but not the 532-nm pump laser. The squeezer is locked on the resonance to allow transmission of the control field to filter cavity. The filter cavity is also locked on resonance with the auxiliary field to mimic the nominal operation with frequency-dependent squeezing. If there is any extra technical noise introduced with frequency-dependent squeezing, for example backscatter noise driven by filter cavity length fluctuations, the interferometer would sense it in the total noise spectra in both configurations. 

To confirm if there are any excessive noises including backscatter, we compare the total unsqueezed interferometer noise with the following two operating modes. The first one is to open the squeezer beam diverter to mimic the frequency-dependent squeezing case as mentioned above, and the second one is to close the squeezer beam diverter on the injection path such that no backscattered light can be transmitted between interferometer and squeezing system. We follow Eq.(13) in \cite{Haocun_nature} to estimate the uncertainties. We have two PSD of each mode and calculate $\delta N(\Omega)$ between PSDs of the same and different operating modes.

\begin{figure}
    \centering
    \includegraphics[width=0.6\linewidth]{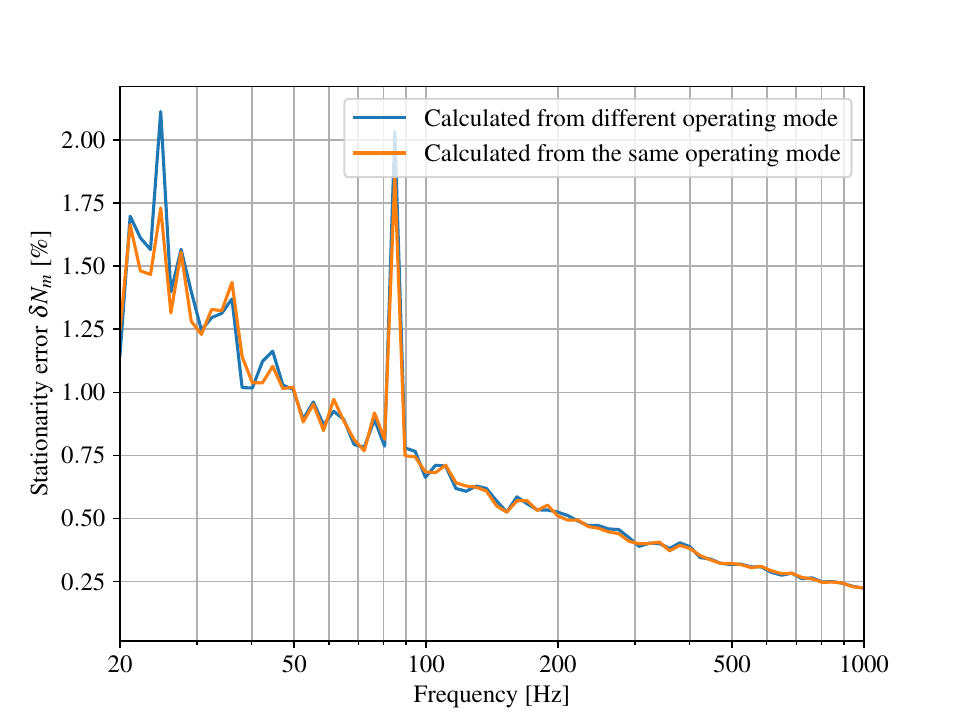}
    \caption{Comparison of the unsqueezed noise stationarity between two PSDs measured in the same unsqueezed operating mode (e.g. two segments with the squeezer beam diverter open), or measured in two different unsqueezed operating modes (e.g. with squeezer beam diverter open and beam diverter closed). Both uncertainties are the same, suggesting the squeezer system does not introduce excess technical noise in the full detectors.
    }
    \label{fig: deltaN_m}
\end{figure}

\cref{fig: deltaN_m} shows that the stationarity uncertainty curves are nearly identical between two PSD taken at the same operating mode or different operating mode, confirming that the mode-nonstationarity contribution to the total stationarity uncertainty is negligible. 

The time-nonstationarity occurs due to thermal drifts of the interferometer. For the faster averaging timescales used in our measurements, slow drifts can be reduced with longer averaging times, similarly to statistical PSD estimation. Therefore, both of drifts and statistical uncertainties are reduced after the re-binning process. 

Calibration uncertainty $\delta G_{cal}(\Omega)$ are estimated in the same way as \cite{Haocun_nature}. Note that it is a form of systematic error instead of statistical error. Therefore, the calibration error is added to the total uncertainty after re-binning, since it can not be reduced by averaging. The contributions of aforementioned uncertainties to the total uncertainty are shown in \cref{fig: errbudget}.

\begin{table}[ht]
\centering
\caption{Parameters of the LIGO Livingston detector inferred using Markov Chain Monte Carlo (MCMC) methods. Fixed and chosen parameters are input parameters for the MCMC, which infers the ``common" and ``independent" parameters. Common parameters are shared across squeeze angle measurements, whereas independent parameters are allowed to change across squeeze angle measurements. See \cite{aLIGO,Aaron} for more detailed parameters of LIGO, including e.g. optic transmissivities.}
\renewcommand{\arraystretch}{1.25}
\begin{tabular}{l c c c c}
    \hline     
    &  \multicolumn{3}{c}{MCMC Set-up} & Inferred  \\
    & Fixed/  & Prior & Initial walker & Common/  \\
    & Chosen  &  Gaussian $(-\sigma, \sigma)$ & Flat probability &  Independent  \\
    \hline
    \textbf{Interferometer parameters}  \\
    \ \ \ \ Circulating power in arm cavity  && (270, 320) kW & 270 - 310 kW & $257_{-1.6}^{+3.9}$ kW \\
    \ \ \ \ Arm to SEC mismatch & 2.7\%  \\
    \ \ \ \ Arm to SEC mismatch phase  & 0$^\circ$  \\
    \ \ \ \ SEC round-trip detuning phase  & $0.14^{\circ}$ \\
    \ \ \ \ SEC round-trip Gouy phase  && (20, 50) $^{\circ}$ & 20$^{\circ}$ - 70$^{\circ}$ & $43.0_{-5.2}^{+4.5} \ ^\circ$  \\
    \ \ \ \ Readout angle &-11$^\circ$\\
    \textbf{Total readout loss } &&(8, 10) \% & 6\% - 10\%& $8.0^{+1.2}_{-0.5}$ \%\\
    \ \ \ \ IFO to OMC mismatch  && (6, 8) \% & 4\% - 10\%& $3.6_{-0.5}^{+0.5}$ \% \\
    \ \ \ \ IFO to OMC mismatch phase  &&&& Independent \\
    \hline
    \textbf{Squeezing parameters}&&\\
    \ \ \ \ Generated squeezing  & 17.4 dB \\
    \ \ \ \ Squeezing angle &  Chosen \\
    \textbf{Total Injection efficiency}  & 92.9\% \\ 
    \ \ \ \ SQZ to OMC mismatch  && (1, 8) \% & 1\% - 8 \% & $1.1_{-0.2}^{+1.3}$ \% \\
    \ \ \ \ SQZ to OMC mismatch phase  & -45$^\circ$ \\
    \textbf{Phase noise} (RMS)  & Chosen  \\
    \hline
    \textbf{Filter cavity parameters}&&\\
    \ \ \ \ Length  & 300 m\\
    \ \ \ \ Detuning    & & ($-28$, $-25$) Hz & $-31$ Hz - $-26$ Hz &  $-25.6$ Hz  \\ 
    \ \ \ \ Finesse  & 7000   \\
    \ \ \ \ Full-linewidth   & 71 Hz   \\
    \ \ \ \ Input coupler transmission && (800, 900) ppm & 750 ppm - 880 ppm & 797 ppm\\
    \ \ \ \ Derived round-trip loss   && && 100 ppm\\
    \ \ \ \ Squeezer to FC mismatch     & 0.2\%\\
    \ \ \ \ Squeezer to FC mismatch phase    && ($-180$, 180) $^{\circ}$ &$-180^{\circ}$ - 180$^{\circ}$&  $-65^\circ$\\
    \ \ \ \ Length noise (RMS)  && (0.1, 1) pm & 0.1 pm - 2 pm & 0.2 pm  \\
    \hline
\end{tabular}
\label{tab: params}
\end{table}

\section{Full Quantum Noise Model}
The only remaining source of uncertainty to be discussed is the modeling uncertainty $\delta M(\Omega)$. In this paper, we use a novel method to estimate and constrain model parameters with Markov Chain Monte Carlo (MCMC) inference. Before discussing details of the inference method, we briefly explain the latest model of quantum noise.

The LIGO detector is essentially an assembly of individual optical cavities. The core optics of the interferometer is composed of two 4-km long arm cavities as two Michelson arms. The two input ports of the Michelson have two partially-reflective mirrors to boost arm power and increase signal bandwidth separately at bright and dark port (see \cref{fig: LIGO}). The squeezing system generates squeezed vacuum using an optical parametric amplifer cavity, performs frequency-dependent rotation with a detuned filter cavity, and couples into the interferometer at the dark port. Each cavity in the system has degredations like optical losses and off-resonance detunings. In addition, there are non-zero mismatches between the fundamental spatial modes of two consecutive cavities. The full model captures all of these non-idealities based on our latest theoretical work \cite{Lee_LIGO_response}.

We use Gravitational Wave Interferometer Noise Calculator (GWINC) to numerically compute the detector quantum noise. It is a phenomenological and analytical model that is derived from input-output relations \cite{buonannoPRD01QuantumNoise}. It extends the optical fields of fundamental (TEM$_{00}$) spatial mode to one higher-order (TEM$_{20}$) mode \cite{buonannoPRD01QuantumNoise, Lee_LIGO_response}. The model includes all the decoherences, degradations, and dephasings of the squeezed vacuum \cite{DnD}. The full sets of parameters can be found in \cref{tab: params}.

For parameter estimation, we rely on external measurements as our priors whenever possible. But, we do not have external measures of several quantities, and for others, external measurements do not have the necessary accuracy and precision, and may vary if not measured in-situ. We use MCMC, informed by external measurements whenever possible, to estimate experimental parameters. The full model of interferometer with frequency-dependent squeezing has a total of 20 impactful parameters. We reduce the problem by isolating the interferometer from the squeezing system first. Then we introduce the squeezer parameters to model frequency-independent squeezing measurements, and finally the filter cavity to model frequency-dependent squeezing measurements.

\subsection{Inferring Interferometer Parameters}

LIGO employs an active calibration system, known as the Photon Calibrator \cite{Pcal}, to calibrate the measured optical power into meters of differential arm length. The system actively modulates the differential arm length by sending an amplitude-modulated laser beam on the test mass. Therefore, we can directly measure the interferometer's transfer function (in units of meters/milliAmp, often called the ``sensing function'') by sweeping the Photon Calibrator laser frequency. 

At the dark port of the interferometer, LIGO has an additional mirror, known as signal recycling mirror with 32.5\% power transmission, to effectively broaden the sensitivity bandwidth to the differential arm length signal. The cavity formed by signal recycling mirror and the interferometer has parameters such as loss, mode-mismatch, and off-resonance detuning, which directly impact the measured sensing function. Therefore, we can isolate and infer these parameters by fitting the sensing function with MCMC. The total parameter space is reduced after we successfully infer parameters of the signal recycling cavity from the sensing function.

\subsection{Inferring Frequency-Independent Squeezing Parameters}

After inferring the parameters of the signal recycling cavity, we feed them into the model that describes the squeezed interferometer. We simplify the squeezing system by bypassing the filter cavity first. It is often difficult to infer model parameters when many parameters of the model are degenerate. For example, mode-mismatch and loss between interferometer and output mode cleaner cavity are degenerate when this mode-mismatch is the only mismatch in the optical path \cite{Lee_LIGO_response}. If we introduce multiple mismatches to break the degeneracy, there are redundant parameters that provide more than one solution to satisfy measurements.

To constrain the quantum noise model, we change the squeezing angle $\phi$ to alter the quantum noise $S_{\text{SQZ}}(\Omega)$ in \cref{eq: model} while keeping the filter cavity end mirror misaligned to simplify the system ($\phi$ is frequency-independent in this case). Since we have only changed the squeezing parameter, there should exist a set of model parameters that can fit all of the measurements by only altering the squeezing angle, if the model fully captures the physics. Assuming such a set of common parameters should break certain degeneracies in the model and constrain the parameter space.

Experimentally, we misalign the filter cavity and change the the squeezing angle $\phi$ by adjusting the offset of locking point of the phase-locking-loop between frequency-independent squeezing and the local oscillator field of the interferometer. We operate the interferometer in an unsqueezed mode (pump laser blocked so no squeezed photons are being generated) and frequency-independently squeezed mode at various $\phi$. 20-minute time series data is taken in each operating mode. Assuming the classical noise $C(\Omega)$ is stationary across different configurations, we can take the difference of two measured total noise PSDs and model the quantum noise differences (\cref{eq: model}),
\begin{equation}
    S_\text{diff}(r,\phi) = D_s(r,\phi) - D_r(r=0) = S(r,\phi) - S(r=0)
\end{equation}
where the total measured noise is $D(\Omega) = S(\Omega) + C(\Omega)$. Although we can not directly measure the classical noise, we can still model the quantum noise difference that is measurable.

\begin{figure}
    \centering
    \includegraphics[width=1.0\linewidth]{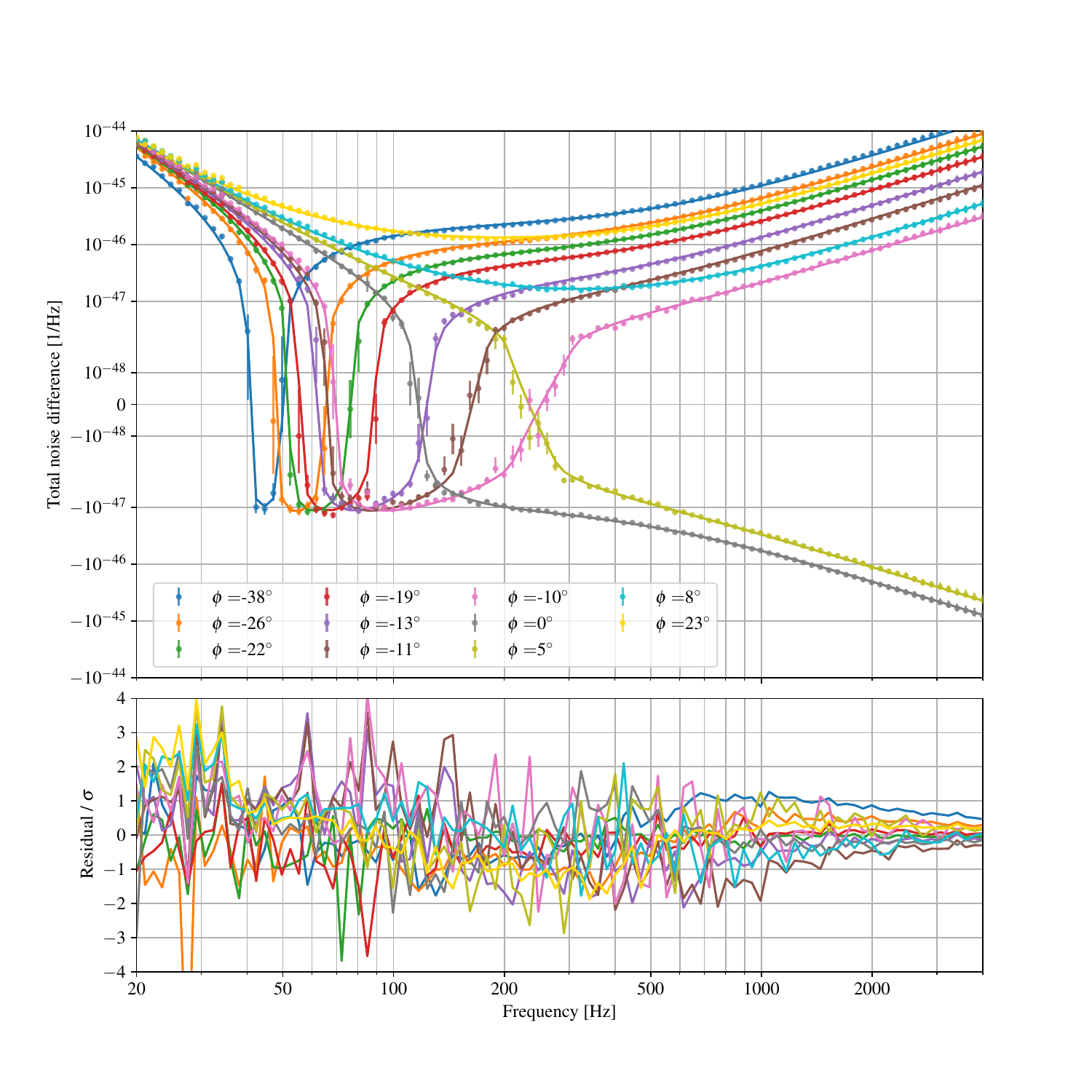}
    \caption{Inference results on the difference of total noise between frequency-independent squeezed and unsqueezed interferometer at various squeezing angles. The negative PSD difference means that the quantum noise is being squeezed. The residual between model and measurements are normalized by the 1-$\sigma$ uncertainty and shown in the bottom plot. }
    \label{fig: MCMC_FIS}
\end{figure}

LIGO reads out optical power fluctuations that are transmitted (cleaned) by the output mode cleaner cavity. The transmitted beam is divided onto two photodetectors using a 50/50 beam splitter, and we read out the summation of the photocurrent signals. The sum reads out the squeezed quantum noise we observe, and the difference of the two photocurrents, known as the null channel, subtracts all of the correlated noise and only leave the uncorrelated noises of the two photodiodes, namely quantum shot noise and dark noise of the detector. The null channel provides a simultaneous monitoring of the calibrated quantum shot noise, which is computed by dividing the flat quantum shot noise in milliAmps by the sensing function. 

We find the best inference of the parameters with MCMC. We use a Gaussian likelihood with a set of Gaussian priors for each parameter. For each measurement with certain squeezing angle, we fit both the noise difference and the quantum shot noise, the latter of which is used to infer the readout loss. The initial walkers are distributed with a flat probability in a bounded interval. As a result, the method is able to find a set of common parameters that minimize the residual of all squeeze angle measurements, as presented in \cref{fig: MCMC_FIS} and \cref{tab: params}.

There are four types of parameters in our inference methods:
\begin{itemize}
    \item ``Fixed" parameters are fixed across all squeeze angle datasets. For example, the signal recycling cavity parameters we inferred earlier are assumed to be the same for all. 
    \item ``Chosen" parameters are selected and different for each squeeze angle dataset. For example, the squeezing angle is actively changed to obtain different squeezing PSD.
    \item ``Common" parameters are shared degrees of freedom that MCMC infers a single value across all squeeze angle datasets. For example, the power within arm cavity should be the same across measurements, and we use MCMC to infer its exact number.
    \item ``Independent" parameters are degrees of freedom of MCMC infers differently for each squeeze angle dataset. 
\end{itemize}

In \cref{tab: params}, we set the squeezing angle and phase noise as ``chosen parameters". It is known that the residual phase noise error of the aforementioned phase-locking-loop depends on the control offset and therefore the squeezing angle. To be able to fit the PSD difference, we still need to set the mode-mismatch phasing between interferometer and output mode cleaner (\cref{fig: LIGO}) as an ``independent parameter", which is an extra phase in the optical path calculated from 2-dimensional overlap integral of the wavefronts of two eigenmodes of two cavities \cite{Lee_LIGO_response}. This mode-mismatch phasing only helps us fit the model phenomenologically, and is not expected to physically depend on the squeezing angle. Instead, the MCMC adjusts this phasing to mimic certain physics that is not fully captured in the latest model in order to fit the measurements. 

\cref{fig: MCMC_FIS} validates our quantum noise model as we successfully fit all of the measurements at various squeezing angles by independently tuning a minimal set of parameters. The model uncertainty $\delta M(\Omega)$ is obtained by taking the 16th and 84th percentile of the model curve computed from the parameters of the MCMC chain (after burning in). Now we collect all sources of uncertainties (\cref{fig: errbudget}) and compute the inferred quantum noise with frequency-dependent squeezing.

\begin{figure}[h!]
    \centering
    \includegraphics[width=0.8\linewidth]{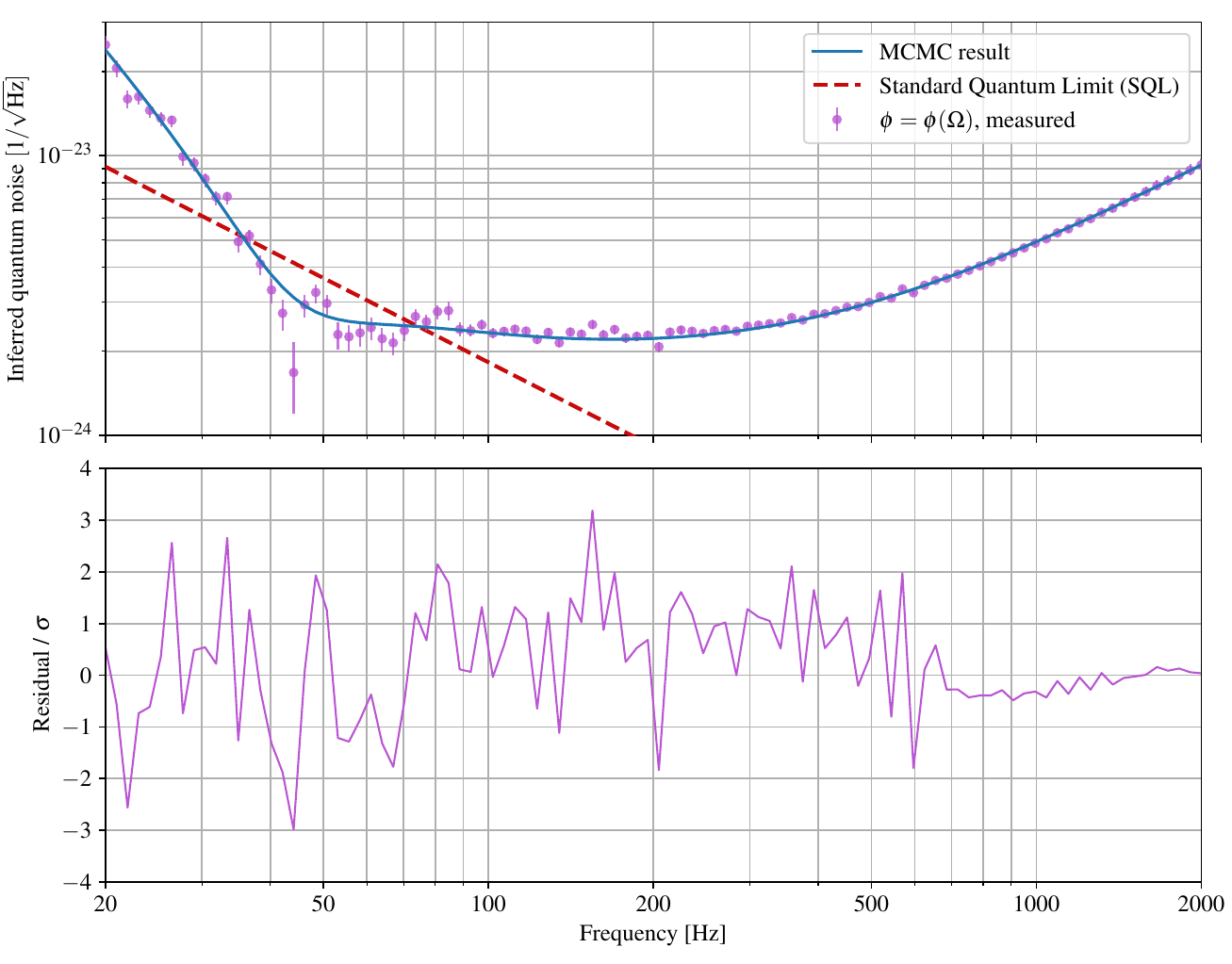}
    \caption{Inference of the quantum noise with frequency-dependent squeezing. }
    \label{fig: MCMC_FDS}
\end{figure}

\subsection{Inferring Filter Cavity Parameters}

Using the interferometer and squeezer quantum noise models obtained in the previous subsections, we can compute the inferred quantum noise ASD with frequency-dependent squeezing from \cref{eq: Q}. We perform a final MCMC to infer the remaining filter cavity parameters.

Since LIGO is currently operating at a lower arm power than the designed value, the filter cavity is not operating in the optimal configuration \cite{whittlePRD20OptimalDetuning}. This is the reason why the current frequency-dependent squeezed quantum noise does not trace the sub-SQL dips of each frequency-independent measurements, in addition to a noise bump near \SI{80}{Hz} due to scattered light. In the MCMC, we assumed the filter cavity finesse to be 7000 in order to fit external cavity ringdown and linewidth measurements of the filter cavity. The inferred parameters are shown in \cref{tab: params}.

\section{Total Uncertainty Budget}
\begin{figure}[h!]
    \centering
    \includegraphics[width=0.8\linewidth]{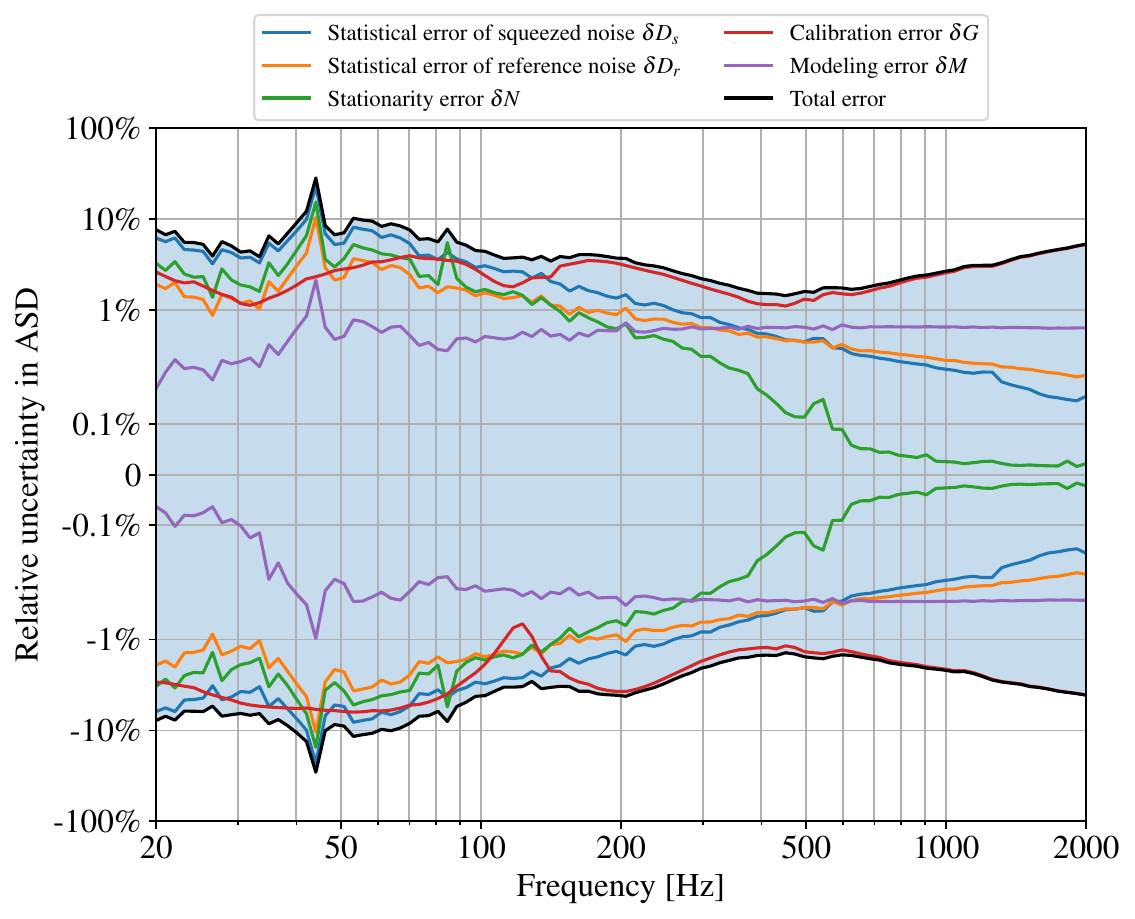}
    \caption{Total uncertainty budget of inferred quantum noise from various error sources. }
    \label{fig: errbudget}
\end{figure}

Now that we have collected all sources of the uncertainties $\delta q (\Omega)$ of the inferred quantum noise amplitude spectral density $q(\Omega)$, we can add these independent noises together in quadrature to obtain the final 1-$\sigma$ uncertainty. The contributions of each uncertainty is shown in \cref{fig: errbudget}.

The statistical uncertainty dominates both positive and negative error bars at low frequency due to the small frequency bin width (\cref{eq: deltaD}). At high frequencies above \SI{500}{Hz}, the statistical error decreases as there are more averages available per bin width. Both statistical and stationarity error are symmetrical, whereas the calibration error and modeling error are not. The calibration error, obtained from the calibration pipeline \cite{calib_paper}, dominates at high frequency above \SI{200}{Hz}. 

Considering all measurement uncertainties, the LIGO detector operates with sub-SQL quantum noise at more than 3-$\sigma$ statistical confidence, as enabled by frequency-dependent squeezing (\cref{fig: moneyplot}).

\section{Sub-SQL Performance}

\begin{figure}[ht]
    \centering
    \includegraphics[width=1.0\linewidth]{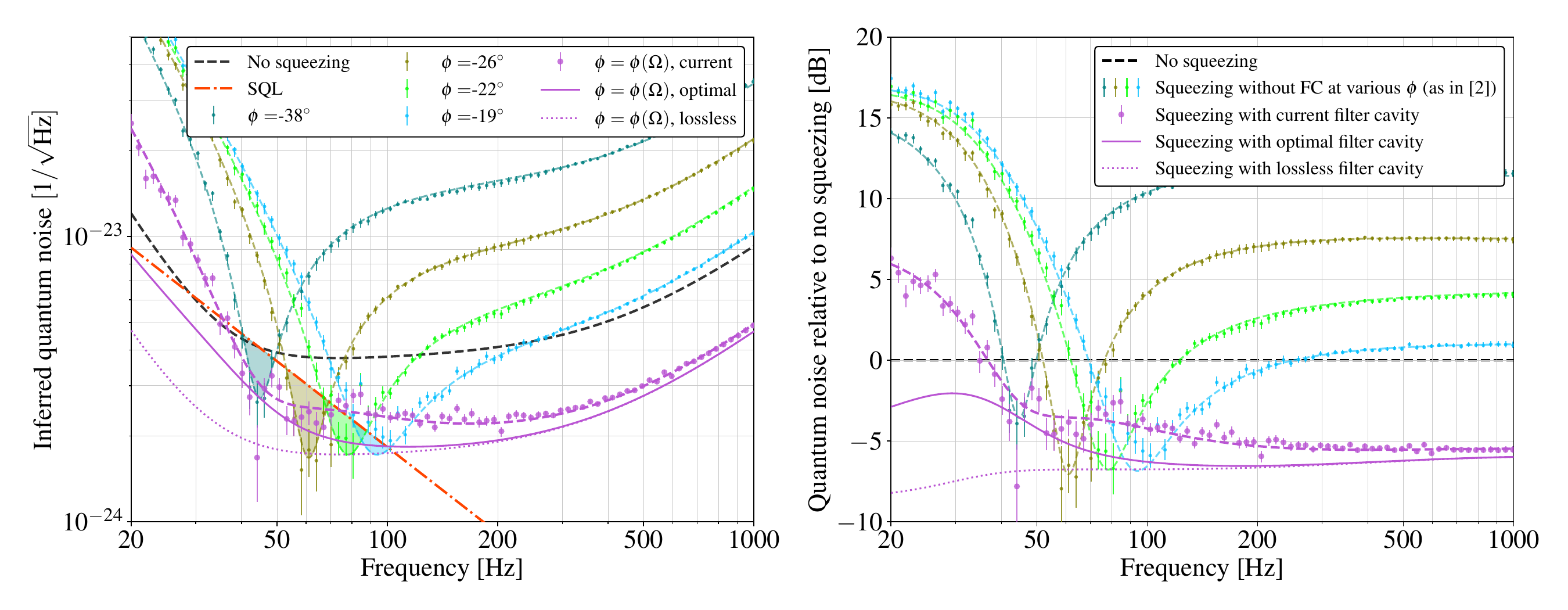}
    \caption{Quantum noise reduction in strain amplitude spectral density. Blue, olive, lime, and teal traces show the inferred quantum noise with frequency-independent squeezing injected at four different squeeze angles $\phi$. The three purple traces show the quantum noise with three frequency-dependent squeezing configurations, same as \cref{fig: QND_FIS}.}
    \label{fig: SQL_FIS}
\end{figure}

\cref{fig: SQL_FIS} compares the sub-SQL performance with frequency-independent squeezing (constant squeezing angle $\phi$) and frequency-dependent squeezing ($\phi = \phi(\Omega)$). The sub-SQL dip can be produced by sending squeezing at a fixed angle, as previously observed \cite{Haocun_nature}. However, the dip has a very narrow frequency range. Although we can move the dip frequency by changing squeezing angle, it is not an optimal configuration for maximum sensitivity at all frequencies. As mentioned in the main text, frequency-dependent squeezing can theoretically achieve the sub-SQL envelope that covers all dips that frequency-independent squeezing can achieve (dotted purple). The current and optimal filter cavity are more realistic configurations, and they are the same as \cref{fig: QND_FIS}.

\section{Future Filter Cavity Upgrade}

\begin{figure}[ht]
    \centering
    \includegraphics[width=0.6\linewidth]{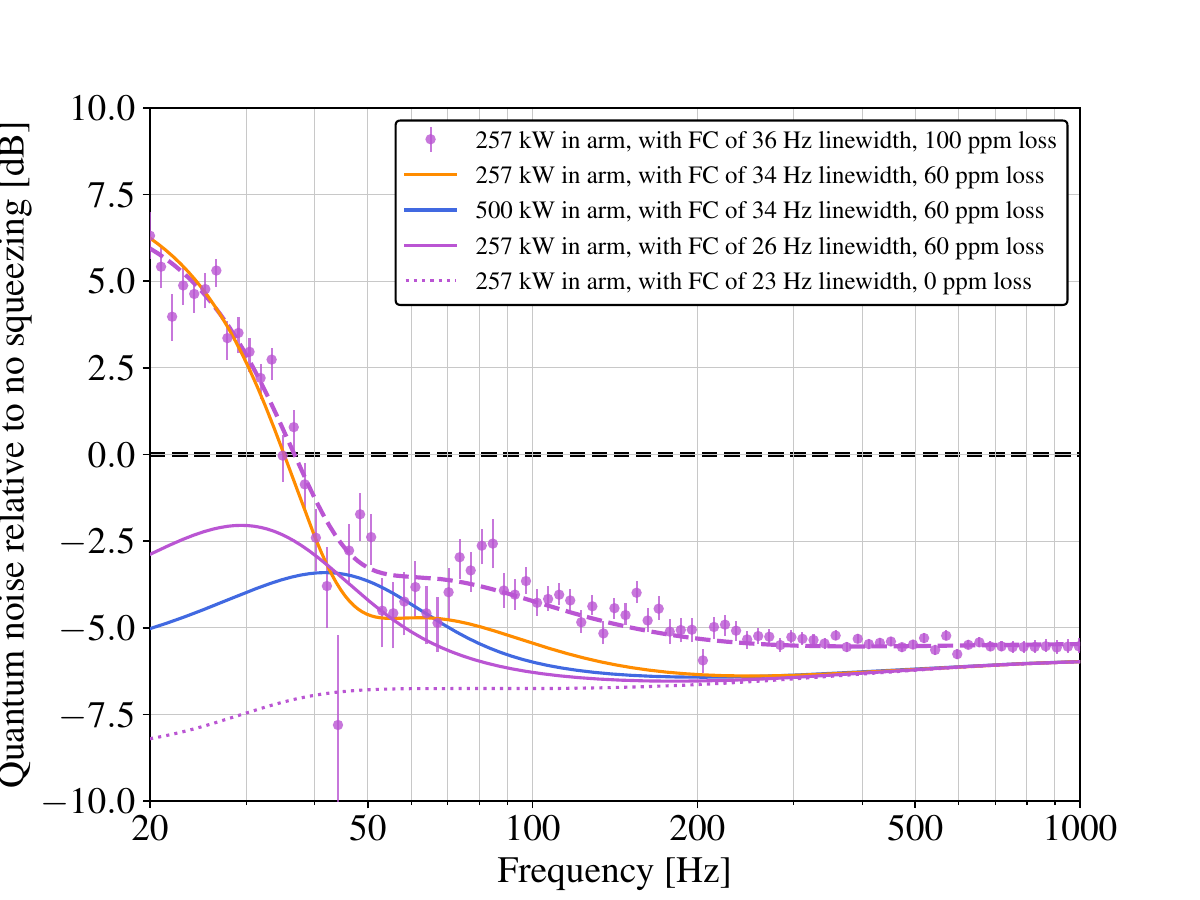}
    \caption{Comparison of the quantum noise with various filter cavity configurations. }
    \label{fig: future_FC}
\end{figure}

While we demonstrate that the optimal lossless filter cavity is able to simultaneously achieve all sub-SQL dips that frequency-independent squeezing can do, we have to acknowledge the fact that a realistic filter cavity has a non-zero loss. The designed round-trip loss of the filter cavity is \SI{60}{ppm}, compared to the loss of \SI{100}{ppm} suggested by our MCMC. A few different filter cavity configurations are shown in \cref{fig: future_FC}.

\newpage

In \cref{fig: future_FC}, the relative quantum noise curves with current filter cavity (dashed purple) and optimal filter cavity (solid purple) are identical to \cref{fig: QND_FIS}. If we achieve the designed loss of 60 ppm with current filter cavity, the squeezing will improve from dashed purple to the orange curve. It is only possible to achieve squeezing at all frequencies when we adjust the filter cavity linewidth $\gamma_\text{FC}$ to approach $\Omega_\text{SQL} / \sqrt{2}$, for example, reducing the filter cavity input coupler transmission to \SI{584}{ppm} (purple curve) or increasing the arm cavity power to \SI{500}{kW} (blue curve). The lossless filter cavity is shown in the dotted purple trace. Note that the squeezing in the lossless case is not flat because we have a nonzero phase difference between the local oscillator field and the signal field, known as the readout angle. For each trace in \cref{fig: future_FC}, the detuning frequency of the filter cavity is optimized to maximize sensitivity to binary neutron star inspirals - a standard figure of merit for gravitational wave detectors. 

\newpage

\begin{figure}[ht]
    \centering
    \includegraphics[width=0.8\linewidth]{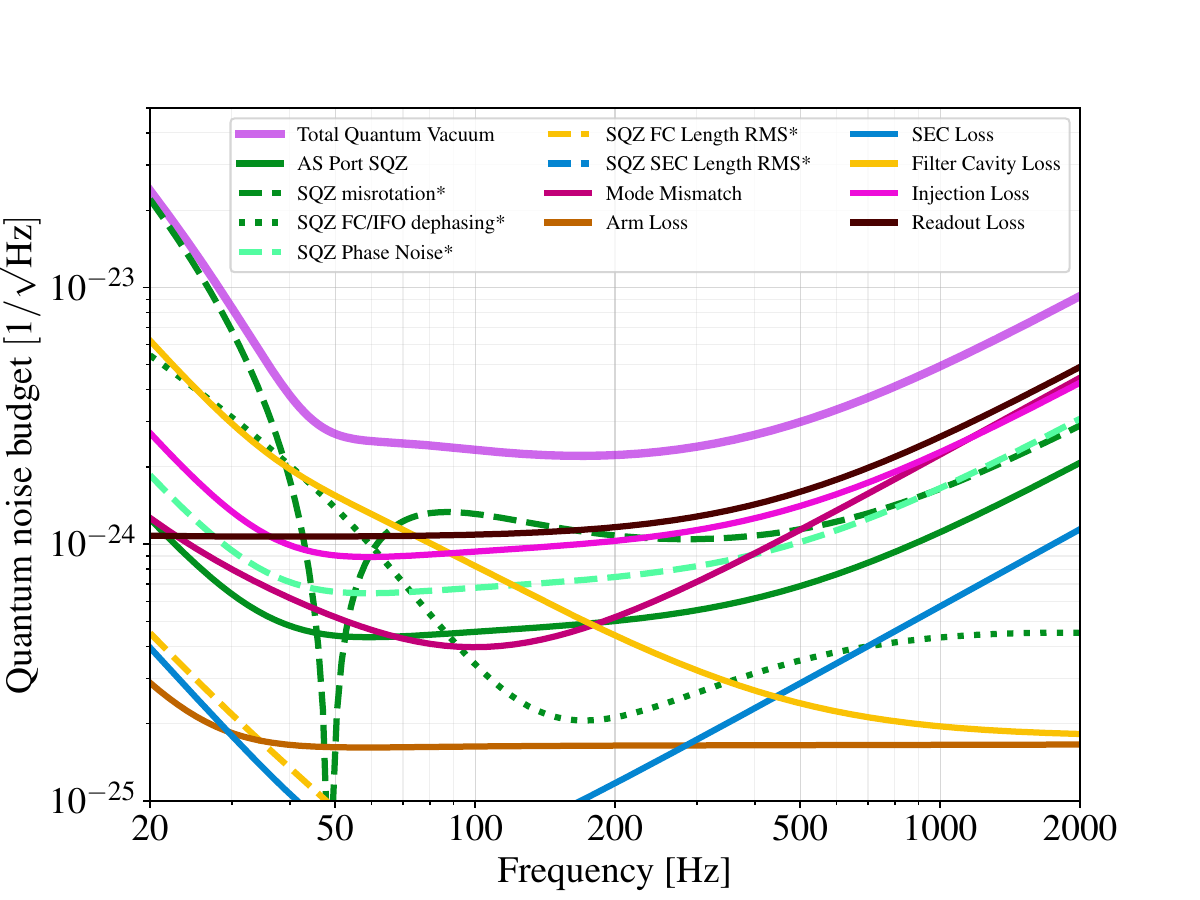}
    \caption{Sub-budget of contributions to the total quantum noise. }
    \label{fig: budget}
\end{figure}

\cref{fig: budget} shows the contributions of the total quantum noise plotted in \cref{fig: moneyplot}. At low frequencies below \SI{40}{Hz}, quantum noise is mostly limited by misrotation of the squeezed state due to the non-optimal filter cavity. At high frequencies above \SI{200}{Hz}, squeezing is limited by the losses due to injection, readout, and mode-mismatches along the optical path. Reducing these major noise sources is the key to further quantum enhancement in the LIGO detectors.

\end{document}